\documentstyle[12pt,aasms4]{article}
\begin{document}

\title {CRITERION FOR GENERATION OF WINDS 
FROM MAGNETIZED ACCRETION DISKS}
\author {OSAMU KABURAKI$^1$}
\affil {$^1$ Astronomical Institute, Graduate School of Science, 
Tohoku University, \\Aoba-ku, Sendai 980-8578, Japan;\\
okabu@astr.tohoku.ac.jp} 

\begin{abstract}

An analytic model is proposed for non-radiating accretion flows accompanied 
by up or down winds in a global magnetic field. 
Physical quantities in this model solution are written in 
variable-separated forms, and their radial parts are simple power law 
functions including one parameter for wind strength. 
Several, mathematically equivalent but physically different expressions 
of the criterion for wind generation are obtained. 
It is suggested also that the generation of wind is a consequence of the 
intervention of some mechanism that redistributes the locally available 
gravitational energy, and that the Bernoulli sum can be a good indicator 
of the existence of such mechanisms. 

\end{abstract}


\section{INTRODUCTION}

Blandford \& Payne (1982) have derived a criterion for the generation 
of centrifugal winds from magnetized accretion disks. 
It has been shown in their self-similar solution that a centrifugal 
wind appears if the inclination of the poloidal magnetic field lines 
penetrating a Keplerian accretion disk makes an angle of less than 
60$^{\circ}$ with the disk surface. 
Although this result is intuitively very understandable, in a more 
general situation that does not satisfy self similarity, this kind of 
criterion becomes meaningless because the inclination angle may be 
different at different locations even along a stream line. 
Therefore, we need to seek for physically more essential expressions 
of the criterion for the generation of winds from accretion disks, in 
order to improve the understanding of such processes (for general 
reviews of the wind and jet theories, see e.g., Begelman, Blandford \& 
Rees 1984; Ferrari 1998; Livio 1999). 

In relation to recent development of the theories of optically thin 
ADAFs (advection-dominated accretion flows: see e.g., Ichimaru 1977; 
Narayan \& Yi 1994, 1995; Abramowicz et al. 1995; Chen et al. 1995), 
the sign of the Bernoulli sum has drawn considerable attention (and 
has caused also confusion) as a possible indicator of the presence 
of unbounded outflows such as winds or jets. 
The sum consists of the gravitational energy, kinetic energy and 
enthalpy, all per unit mass of a fluid. 

Narayan \& Yi (1994, 1995) have shown that the sign of the Bernoulli 
sum is necessarily positive in their self-similar ADAF solution, and 
argued that this is a genuine property of general ADAFs. 
They have suggested also that a flow with positive Bernoulli sum 
(and hence ADAFs, specifically) can easily drive winds and jets. 
Further developed and popularized by Blandford \& Begelman (1999) in 
their influential ``ADIOS" (adiabatic inflow-outflow solution) 
paper, these suggestions have been widely accepted. 
The latter paper stresses and demonstrates the necessity for including 
winds or jets in constructing a satisfactory theoretical model of 
inefficiently radiating accretion flows. 

Meanwhile, several authors have demonstrated that the positivity of 
the Bernoulli sum is not a genuine property of ADAFs. 
As shown by Nakamura (1998), the analytic solution of Honma (1996) 
that includes a diffusion cooling by turbulence has negative sign 
at least in a special case of $\gamma=5/3$, where $\gamma$ is the 
index of polytrope. 
Further it has been shown that, under the influence of convective 
cooling (Igumenshchev \& Abramowicz 2000) and of the inner and outer 
boundary conditions (Abramowicz, Lasota, \& Igumenshchev 2000), 
low viscosity ADAFs (with $\alpha<0.3$, where $\alpha$ is the viscosity 
parameter) have negative sums. 
This result agrees with all of the subsequent 1-D, 2-D and 3-D 
numerical simulations by Abramowicz and his collaborators. 
For inviscid ADAFs, the no-wind solution in a global magnetic field 
(Kaburaki 2000, hereafter K00) shows that the Bernoulli sum is always 
zero within the adopted approximation. 

Another point to be mentioned is the claim by Abramowicz, Lasota, \& 
Igumenshchev (2000) that a positive Bernoulli sum is only a necessary 
but not the sufficient condition for unbounded outflows, as shown 
for inviscid, non-magnetic fluids. 
They also argue that the best example in which a positive Bernoulli sum 
does not imply unbounded outflows is the classical Bondi solution for 
spherical accretion flows. 

To summarize the present status of the Bernoulli sum described above, 
at least, the positivity of the sum does not seem to be a genuine 
property of ADAFs. 
The sign may depend on various conditions such as dissipation 
mechanisms (viscous or resistive), the presence of energy transport
in a fluid (convection, conduction etc.) and the effects of boundaries. 
However, it is still controversial whether the sign of the Bernoulli 
sum can be a direct indicator of the presence of unbounded outflows. 

It seems natural to believe that the sum is everywhere zero (the 
asymptotic value at infinity) in a fluid, as far as the radiation 
cooling and other mechanisms of energy redistribution in the fluid 
are completely negligible, since then all the dissipated energy 
remains as thermal energy of each fluid element and is merely 
advected down the flow (i.e., the flow is completely advective). 
Therefore, a non-zero Bernoulli sum should indicate the intervention 
of some mechanism of energy redistribution within a non-radiative 
flow. 
We may expect convection, conduction or fluid viscosity as such 
mechanisms of redistribution. 

In the present paper, we will derive an analytic model that describes 
non-radiating accretion flows accompanied by up or down winds in a 
global magnetic field, based on our previous treatment, K00. 
Although the main motivation of the present work is the desire to 
improve broadband spectral fittings to low-luminosity active galactic 
nuclei (LLAGNs) and normal galaxies (e.g. for Sgr A$^*$, see Kino, 
Kaburaki \& Yamazaki 2000; Yamazaki, Kaburaki \& Kino 2001), the 
resulting solution will be useful to discuss the issues about the 
Bernoulli sum discussed above.

\section{SIMPLIFYING ASSUMPTIONS}

A schematic drawing of the global configuration presumed in the present 
consideration is given in Figure 1 of K00. 
The viscosity of accreting plasma is completely neglected to clarify the 
role of magnetic field. 
An asymptotically uniform magnetic field is vertically penetrating 
the accretion disk and is twisted by the rotational motion of the plasma. 
Owing to the Maxwell stress of this twisted magnetic field, a certain 
fraction of the angular momentum of accreting plasma is carried out 
to infinity, and this fact enable the plasma gradually infall toward 
the central black hole. 

As shown in Appendix of K00, the geometry of accretion flows is 
very essential in specifying their physical properties. 
For example, the ADAF solution follows almost automatically from 
the assumption of constant opening angle of the disk (i.e., 
$\Delta=$ const., where $\Delta$ is the half-opening angle), which 
respects the spherical nature of the gravity. 
In this sense, spherical polar coordinates ($r$, $\theta$, $\varphi$) 
are convenient for the discussion of ADAFs. 
For the relevant physical quantities, we follow the notation of K00 
unless specified otherwise. 

Further simplifying assumptions made in K00 in obtaining the no-wind 
ADAF solution from the set of resistive MHD equations were those of, 
i) stationarity ($\partial/\partial t = 0$), ii) axisymmetry 
($\partial/\partial{\varphi} = 0$), iii) geometrically thin disk, 
iv) weakly resistive disk, v) dominance of midplane, and vi) no wind. 
Among these assumptions, only the last one will be removed in the 
present paper in order to obtain the ADAF solution including winds.

The third assumption means that $\Delta\ll 1$, and is most effective 
to simplify the basic equations. 
It has been shown in K00 that the presence of an external magnetic 
field ${\bf B}_0$ guarantees the realization of such a thin disk even 
in a hot ADAF situation. 
Reflecting this localized structure, we introduce an angular variable 
$\eta = (\theta-\pi/2)/\Delta$. 
Then it becomes clear that a differentiation with respect to $\theta$ 
gives rise to a quantity of order $\Delta^{-1}$ 
($\partial/\partial\theta = \Delta^{-1}\partial/\partial\eta$). 
We can also safely approximate in the disk as $\sin\theta \simeq 1$ and 
$\cos \theta \simeq 0$. 

The statement that a disk is weakly resistive implies that the 
``characteristic" magnetic Reynolds number $\Re$, whose definition will 
be introduced later, is large in the sense that $\Re^2(r) \gg 1$ in the 
disk except near its inner edge $r_{\rm in}$ where 
$\Re(r_{\rm in}) = 1$. 
Actually, the terms of ${\cal O}(\Re^{-2})$ were neglected in K00. 
In these situations, the externally given magnetic field ${\bf B}_0$ 
is largely stretched by the rotational and infalling motion of the 
accreting plasma so that the deformation in the disk becomes much larger 
than the seed field (if we divide the total magnetic field in the form 
${\bf B} = {\bf B_0} + {\bf b}$, then $\vert{\bf b}\vert\gg
\vert{\bf B_0}\vert$). 
The definition of the disk's outer edge is the radius within which the 
deformation of the poloidal magnetic field becomes noticeable 
(i.e., $\vert{\bf b}_{\rm P}\vert \sim \vert{\bf B}_0\vert$).

The assumption v) naturally follows from the consideration that, 
since majority of matter is concentrated around the midplane of the 
accretion disk, its physical properties should be controlled mainly by 
this part of the disk. 
Therefore, we may seek the approximate solution that is accurate near 
the midplane even if making a sacrifice of the accuracy at its upper 
and lower surfaces. 
According to this spirit, we ignore the quantities that are proportional 
to $\tanh^2 \eta$ since $\tanh^2\eta \ll 1$, $\mbox{sech}^2\eta$. 

The last condition in the above list is, in fact, not indispensable in 
obtaining a resistive ADAF solution in a global magnetic field (or 
resistive ADAF solution). 
Indeed, an accretion flow accompanied by a converging flow (or down wind) 
toward the disk surface has been obtained by the present author 
(Kaburaki 1987) omitting this assumption. 
It was imposed again in K00, however, in order to understand the energy 
budget clearly and to firmly establish a basic analytic model in the 
resistive ADAF regime. 
This condition actually consists of two equations, $v_{\theta} = 0$, and 
$j_{\theta} = 0$ for consistency. 
The former results in an $r$-independent mass accretion rate $\dot{M}$, 
and the latter specifies the radial dependence of the toroidal magnetic 
field as $b_{\varphi}\propto 1/r$ since $j_{\theta}=-(c/4\pi r)\ 
\partial(rb_{\varphi})/\partial r$ under the assumption of ii). 

In relation to the no-wind resistive ADAF solution obtained in K00, it 
should be stressed that the result actually contains the effects of finite 
resistivity to the first order in the smallness parameter $\Re^{-1}$. 
The ratio of poloidal to toroidal magnetic field is small 
(i.e., $b_r/b_{\varphi}\sim\Re^{-1}$) reflecting the strong 
twisting of the poloidal seed field by the rotational motion. 
This ratio is maintained by the balance between this twisting and 
untwisting by resistive diffusion. 
The thin disk structure of the flow is maintained by the vertical 
force balance between the magnetic pressure of $b_{\varphi}$ toward 
the equatorial plane and the opposing gas pressure in the disk. 
Reflecting this plasma enhancement in the disk, the gas pressure and 
density in the disk are the quantities of order $\Re^2$. 
The finite thickness of the disk itself is also a consequence of 
non-zero resistivity (indeed, $\Delta\propto\Re^{-1}$). 
The infall velocity is small in the sense that $v_r/v_{\varphi}
\sim\Re^{-1}$. 
Since the magnetic Reynolds number is a function of $r$, the solution 
does not have self similarity (i.e., in addition to the presence of angular 
dependences, both the ratios $b_r/b_{\varphi}$ and $v_r/v_{\varphi}$ vary 
with the radial distance).

\section{REMOVAL OF NO-WIND CONDITION}
In order to obtain a more general form of the resistive ADAF solution, 
we remove here the no-wind condition from our list of simplifying 
assumptions. 
Fortunately, however, it turns out that all but one terms can be omitted 
finally among the newly appeared terms in the leading order equations in 
$\Delta$, by the aid of the assumptions iv) and v). 
We shall check this point below, expecting that the angular dependence 
of every quantity (see K00, and \S 4 below) is not affected by the 
removal of no-wind condition. 
 
The component expressions of the resistive MHD equations simplified 
under the assumptions i) through iii) are as follows. 
They are correct to the leading order in the powers of $\Delta$. 
In deriving these equations, all quantities except $b_{\theta}$ and 
$v_{\theta}$ (which are of the order of $\Delta$ as confirmed by 
equations [\ref{eqn:mcns}] and [\ref{eqn:fcns}]) are regarded as of 
order unity with respect to $\Delta$. 
\begin{itemize}
  \item {\bf equation of motion}\\
      $r$-component
     \begin{eqnarray}
      \lefteqn{\left( v_r\frac{\partial}{\partial r} 
         + \frac{v_{\theta}}{\Delta r}\frac{\partial}{\partial \eta} 
         \right)v_r - \frac{v_{\varphi}^2}{r}} \nonumber \\
       & & = -\frac{1}{\rho}\ \frac{\partial p}{\partial r} 
       -\frac{GM}{r^2} + \frac{1}{4\pi\rho r} \left[ 
       \frac{b_{\theta}}{\Delta}\ \frac{\partial b_r}{\partial \eta} 
       -b_{\varphi}\frac{\partial}{\partial r}(rb_{\varphi}) \right] 
     \end{eqnarray}
      $\theta$-component
     \begin{equation} 
        p + \frac{1}{8\pi}\left(b_r^2+b_{\varphi}^2\right)=\tilde{p}(r) 
     \label{eqn:EOMth}
     \end{equation}
      $\varphi$-component
     \begin{eqnarray}
        \lefteqn{\left(v_r\frac{\partial}{\partial r}
        +\frac{v_{\theta}}{\Delta r}\ \frac{\partial}{\partial\eta}\right)
        v_{\varphi} + \frac{v_{\varphi}v_r}{r}} \nonumber \\
      & & = \frac{1}{4\pi\rho r}\left[ b_r\frac{\partial}
        {\partial r}(rb_{\varphi})
        +\frac{b_{\theta}}{\Delta}\ \frac{\partial b_{\varphi}}{\partial\eta}
        \right] 
     \end{eqnarray}
  \item {\bf induction equation}\\
      poloidal component
     \begin{equation}
        \frac{1}{c}\ (v_r b_{\theta}-v_{\theta} b_r) 
        = -\frac{c}{4\pi\sigma\Delta}\ \frac{1}{r}\ 
        \frac{\partial b_r}{\partial\eta}
     \end{equation}
      $\varphi$-component
     \begin{equation}
       r^2 b_r \frac{\partial}{\partial r} 
       \left(\frac{v_{\varphi}}{r}\right) 
       - \frac{\partial}{\partial r}(rv_r b_{\varphi}) 
       + \frac{c^2}{4\pi\sigma\Delta^2 r}
       \ \frac{\partial^2 b_{\varphi}}{\partial\eta^2} = 0
     \label{eqn:indph}
     \end{equation}
  \item {\bf mass continuity}
     \begin{equation}
        \frac{1}{r^2}\ \frac{\partial}{\partial r}(r^2\rho v_r) 
        + \frac{1}{\Delta r}\ \frac{\partial}{\partial\eta}(\rho v_{\theta})
        = 0
     \label{eqn:mcns}
     \end{equation}
  \item {\bf magnetic flux conservation}
     \begin{equation}
        \frac{1}{r^2}\ \frac{\partial}{\partial r}(r^2 b_r) 
        + \frac{1}{\Delta r}\ \frac{\partial b_{\theta}}{\partial\eta}
        = 0
     \label{eqn:fcns}
     \end{equation}
\end{itemize}

In $r$-component of the equation of motion, there appear two new 
terms reflecting the removal of the no wind condition. 
They are the second term on the left-hand side and the last term 
on the right. 
Both of them are dropped, however, from this equation because of the 
assumption v) since they are proportional to $\tanh^2\eta$. 
Further, the first term on the left and the third term on the right are 
dropped, as in the no-wind case, owing to the assumption iv) since 
$v_r \propto \Re^{-1}$ and $\rho\propto\Re^{-2}$. 
Thus, the equation becomes finally so simple as 
  \begin{equation} 
         \frac{GM}{r^2}=\frac{v_{\varphi}^2}{r}
         -\frac{1}{\rho}\ \frac{\partial p}{\partial r}. 
  \label{eqn:EOMr}
  \end{equation} 

There is no new term in equation (\ref{eqn:EOMth}), and the second term 
on the left is neglected owing to the assumption iv).
The resulting equation describes the magnetic confinement of the disk 
plasma by $b_{\varphi}$. 
In $\varphi$-component of the equation of motion, the second term 
on the left and the first term on the right contain $v_{\theta}$ and 
$j_{\theta}$, respectively. 
However, the former vanishes since $v_{\varphi}$ is independent of $\theta$ 
and the latter drops because of $\tanh^2\eta$ dependence. 
The resulting form is, as before, 
\begin{equation}
  v_r\frac{\partial(rv_{\varphi})}{\partial r} = \frac{b_{\theta}}
   {4\pi\rho\Delta}\ \frac{\partial b_{\varphi}}{\partial\eta}
  \label{eqn:EOMph}.
\end{equation}

The poloidal components of the magnetic induction equation are degenerate 
reflecting the degeneracy in Maxwell's equations. 
The second term on the left of this equation can be dropped again 
by the assumption v), and we have 
\begin{equation}
  v_r b_{\theta} = -\frac{c^2}{4\pi\sigma\Delta}\ \frac{1}{r}
      \frac{\partial b_r}{\partial \eta}
  \label{eqn:indpl}
\end{equation}
There appears no change in equation (\ref{eqn:indph}). 

Thus, the only change that cannot be dropped in the set of basic 
equations describing magnetized ADAFs including winds is the second term 
on the left-hand side of equation (\ref{eqn:mcns}), which results in a 
radius-dependent mass accretion rate. 
The equation of magnetic flux conservation is the same as before.

\section{SEPARATION OF VARIABLES}

As written out in K00, the the set of resistive MHD equations ([1] $\sim$ 
[4] there) are 8 equations for 8 unknowns (i.e., for $\rho$, $p$, {\bf v} 
and {\bf B}), and hence the set is closed apparently. 
Other quantities such as {\bf j}, {\bf E}, $q$ (the charge density), 
and $T$ (the temperature) are calculated from the subsidiary equations, 
(5) and (6) in K00. 
Actually, however, the above main set is not closed because of the 
degeneracy in Maxwell's equations. 
Usually, this point is supplemented by adding one more equation relating 
the transfer of energy. 

Although, in many cases, the polytropic relation is adopted for simplicity 
as such an equation, its validity is rather doubtful except in the special 
cases of adiabatic and isothermal processes. 
On the other hand, the full equation for the energy transfer, not only 
in the fluid but also to the surroundings by the radiation, is too 
complicated to be treated analytically. 
Here, we would rather let the set being open as in K00 since an open set 
does not mean the absence of solutions. 
Actually, it means only that there is no systematic way of solving it. 

Instead of adding energy equation, we have put two specific constraints 
in K00 in order to obtain the ADAF solution without wind. 
Before entering into the discussion of them, we first separate the 
variables in the following form being led by the experience in K00. 
  \begin{equation}
     b_r(\xi, \eta) =\ \tilde{b}_r(\xi)\ \mbox{sech}^2\eta \tanh\eta, 
  \end{equation}
  \begin{equation}
     b_{\theta}(\xi, \eta) =\ \tilde{b}_{\theta}(\xi)\ \mbox{sech}^2\eta, 
  \end{equation}
  \begin{equation}
     b_{\varphi}(\xi, \eta) =-\tilde{b}_{\varphi}(\xi)\tanh\eta, 
  \end{equation}
  \begin{equation}
     v_r(\xi, \eta) =-\tilde{v}_r(\xi)\ \mbox{sech}^2\eta, 
  \end{equation}
  \begin{equation}
     v_{\theta}(\xi, \eta) = \tilde{v}_{\theta}(\xi)\tanh\eta, 
  \label{eqn:vth}
  \end{equation}
  \begin{equation}
     v_{\varphi}(\xi, \eta) = \tilde{v}_{\varphi}(\xi), 
  \end{equation}
  \begin{equation}
     p(\xi, \eta) =\ \tilde{p}(\xi)\ \mbox{sech}^2\eta, 
  \end{equation}
  \begin{equation}
     \rho(\xi, \eta) =\ \tilde{\rho}(\xi)\ \mbox{sech}^2\eta, 
  \end{equation}
  \begin{equation}
     T(\xi, \eta) = \tilde{T}(\xi), 
  \end{equation}
  \begin{equation}
     j_r(\xi, \eta) =-\tilde{j}_r(\xi)\ \mbox{sech}^2\eta, 
  \end{equation}
  \begin{equation}
     j_{\theta}(\xi, \eta) = \tilde{j}_{\theta}(\xi)\tanh\eta, 
  \label{eqn:jth}
  \end{equation}
  \begin{equation}
     j_{\varphi}(\xi, \eta) =-\tilde{j}_{\varphi}(\xi)\ \mbox{sech}^4\eta, 
  \end{equation}
  \begin{equation}
     E_r(\xi, \eta) = \tilde{E}_r(\xi)\ \mbox{sech}^2\eta, 
  \end{equation}
  \begin{equation}
     E_{\theta}(\xi, \eta) = \tilde{E}_{\theta}(\xi)\ \mbox{sech}^2\eta
      \tanh\eta, 
  \end{equation}
  \begin{equation}
     E_{\varphi}(\xi, \eta) = \tilde{E}_{\varphi}(\xi)\ \mbox{sech}^4\eta, 
  \end{equation}
where the radial coordinate is normalized by a reference radius $r_0$ as 
$\xi=r/r_0$. 
In the problems of accretion in a asymptotically uniform magnetic field, 
it is natural to choose $r_{\rm out}$ as $r_0$. 
The sign of $\tilde{v}_{\theta}(\xi, \eta)$ is chosen for positive 
$\tilde{v}_{\theta}$ to correspond to an up wind (outflow) from the 
disk surfaces. 

The set of basic equations are now rewritten as the set of ordinary 
differential equations for the radial part functions:
\begin{equation}
  \frac{\tilde{v}_{\varphi}^2}{r} = \frac{1}{\tilde{\rho}}\ 
    \frac{d\tilde{p}}{dr} + \frac{GM}{r^2},
  \label{eqn:Reomr}
\end{equation}
\begin{equation}
  \tilde{p} = \frac{\tilde{b}_{\varphi}^2}{8\pi},
  \label{eqn:Reomth}
\end{equation}
\begin{equation}
  \tilde{v}_r\frac{dl}{dr} = \frac{1}{4\pi\Delta}\ 
    \frac{\tilde{b}_{\theta}\tilde{b}_{\varphi}}{\tilde{\rho}},
  \label{eqn:Reomph}
\end{equation}
\begin{equation}
  \tilde{v}_r\tilde{b}_{\theta} = \frac{c^2}{4\pi\sigma\Delta}\ 
    \frac{\tilde{b}_r}{r},
  \label{eqn:Rindp}
\end{equation}
\begin{equation}
  \tilde{b}_r \left[ r^2\frac{d\Omega}{dr} \right] 
    - \frac{d}{dr}(r\tilde{v}_r\tilde{b}_{\varphi}) 
    + \frac{c^2}{2\pi\sigma\Delta^2}\ \frac{\tilde{b}_{\varphi}}{r}
    = 0,
  \label{eqn:Rindph}
\end{equation}
\begin{equation}
  \frac{\tilde{v}_{\theta}}{\tilde{v}_r} = \Delta \left[ r\frac{d}{dr}
    \ln(r^2\tilde{\rho}\tilde{v}_r) \right], 
  \label{eqn:Rmcns}
\end{equation}
\begin{equation}
  \frac{\tilde{b}_{\theta}}{\tilde{b}_r} = \frac{\Delta}{2} \left[ 
    r\frac{d}{dr}\ln(r^2\tilde{b}_r) \right],
  \label{eqn:Rfcns}
\end{equation}
where we have defined $l\equiv r\tilde{v}_{\varphi}$ and $\Omega 
\equiv \tilde{v}_{\varphi}/r$. 
The angular dependences of the relevant quantities are mutually 
consistent within the assumption v). 

In addition to the above set, the subsidiary equations are 
\begin{equation}
  \tilde{T} = \frac{\bar{\mu}}{R}\ \frac{\tilde{p}}{\tilde{\rho}},
  \label{eqn:RT}
\end{equation}
\begin{equation}
  \tilde{j}_r = \frac{c}{4\pi\Delta}\ \frac{\tilde{b}_{\varphi}}{r},
  \label{eqn:Rjr}
\end{equation}
\begin{equation}
  \tilde{j}_{\theta} = \frac{c}{4\pi}\ \frac{1}{r}\frac{d}{dr}
    (r\tilde{b}_{\varphi}),
  \label{eqn:Rjth}
\end{equation}
\begin{equation}
  \tilde{j}_{\varphi} = \frac{c}{4\pi\Delta}\ \frac{\tilde{b}_r}{r},
  \label{eqn:Rjph}
\end{equation}
\begin{equation}
  \tilde{E}_r = -\frac{\Delta}{2}\ \frac{d}{dr}(r\tilde{E}_{\theta}), 
\end{equation}
\begin{equation}
  \tilde{E}_{\theta} = \frac{1}{c}\ (\tilde{v}_r\tilde{b}_{\varphi} 
    - \tilde{v}_{\varphi}\tilde{b}_r),
\end{equation}
\begin{equation}
  \tilde{E}_{\varphi} = \frac{1}{c}\ \left( \tilde{v}_r\tilde{b}_{\theta} 
    - \frac{c^2}{4\pi\sigma\Delta}\ \frac{\tilde{b}_r}{r} \right) = 0, 
\end{equation}
where $R$ is the gas constant, $\bar{\mu}$ is the mean molecular weight, 
and $\tilde{E}_{\varphi}$ should vanish owing to the assumption ii).

\section{IRAF CONDITIONS}

As mentioned in the previous section, two specific constraints have been 
placed in obtaining the solution in K00. 
Then, the solution has been retrospectively shown to be fully advective. 
Thus, the constraints have taken the place of energy equation, and 
selected the ADAF solution among others. 
Therefore, they may be called the ADAF conditions. 

We shall adopt the same constraints also in the present paper in order to 
select the corresponding specific kind of solution to the above set of 
resistive MHD equations. 
In view of the resulting solution, however, it seems to be needed to 
extend the concept of ADAF here. 
The essence of the conventional ADAF is in that the flow is a very 
ineffective emitter of radiation field, rather than in the point that 
internal energy is transported mainly in the form of advection. 
Such a character is therefore better specified by the term, 
inefficiently-radiating accretion flow or ``IRAF". 
In such a flow, thermal energy may in general be transported by 
convection or by conduction as well as by advection. 
Therefore, we call hereafter the specific constraints the IRAF conditions. 

The first of them is 
  \begin{eqnarray}
  \alpha &\equiv& -\frac{1}{\rho}
    \frac{\partial p}{\partial r}\biggm/\frac{GM}{r^2} \nonumber \\
    &=& -\frac{1}{\tilde{\rho}}\frac{d\tilde{p}}{dr}\biggm/\frac{GM}{r^2}
    = {\rm const.}\ (<1).
  \label{eqn:IRAF1}
  \end{eqnarray}
Then, equation (\ref{eqn:EOMr}) or (\ref{eqn:Reomr}) results in a reduced 
Keplerian rotation of the form 
  \begin{equation} 
        v_{\varphi}=(1-\alpha)^{1/2}\ v_{\rm K}(r), 
  \label{eqn:redK}
  \end{equation}
where $v_{\rm K}(r)\equiv(GM/r)^{1/2}$ is the Kepler velocity. 
The first IRAF condition requires that the pressure gradient term, which 
appears in the equation of radial force balance (\ref{eqn:EOMr}) together 
with the centrifugal force density, behaves exactly in the same way as the 
gravitational force density (i.e., they have the same $r$- and 
$\theta$-dependences). 
In general, however, the former is determined as a complicated 
consequence of the thermal processes such as dissipative heating, 
radiative cooling and advective cooling, and hence may have different 
$r$- and $\theta$-dependences from the latter. 
In spite of this affair, equation (\ref{eqn:IRAF1}) demands the 
pressure gradient to adjust itself to follow the form of the gravity. 
Here, one can see the dominance of the gravity over the thermal 
processes in the IRAF solution. 

The second IRAF condition requires that 
\begin{eqnarray}
   \beta &\equiv& \frac{1}{b_r}\ \frac{\partial}{\partial r}
    (rv_r b_{\varphi}) \biggm/ r^2\frac{\partial}{\partial r} 
    \left(\frac{v_{\varphi}}{r}\right) \nonumber \\
   &=& \frac{1}{\tilde{b}_r}\ \frac{d}{dr}(r\tilde{v}_r\tilde{b}_{\varphi})
    \biggm/ r^2\ \frac{d\Omega}{dr} = \mbox{const.}\ (<1) 
  \label{eqn:IRAF2}
\end{eqnarray} 
Similarly to the first condition, the gravity that specifies the rotation 
law in the denominator determines the degree of magnetic field twist 
as expressed by the numerator. 
In general, however, the latter is determined by a complicated 
consequence of the electrodynamic processes and hence may have different 
$r$-, and $\theta$-dependences from the former. 
Here, we see the dominance of the gravity again, this time, over the 
electrodynamic processes. 

Then, equation (\ref{eqn:indph}), or (\ref{eqn:Rindph}) reduces to 
\begin{equation} 
   \tilde{b}_r  = -\frac{c^2}{4\pi(1-\beta)\sigma\Delta^2} 
     \left[ r^3 \frac{d\Omega}{dr} \right]^{-1} \tilde{b}_{\varphi}. 
\end{equation}
Substituting equation (\ref{eqn:redK}) for $\Omega$, we obtain 
\begin{equation}
  \frac{\tilde{b}_{\varphi}}{\tilde{b}_r} 
    = \frac{3\pi(1-\beta)\Delta^2\sigma}{c^2}\ (1-\alpha)^{1/2}l_{\rm K} 
    \equiv \Re(r),
  \label{eqn:tRe}
\end{equation}
where $l_{\rm K}\equiv\sqrt{GMr}$. 
This ratio, which is specified by the reduced-Keplerian angular-momentum 
distribution, is adopted as the characteristic magnetic-Reynolds number 
of this solution. 
It must be noted in this regard that the actual magnetic Reynolds number 
that is defined by the ratio of the convection to diffusion terms in 
the magnetic induction equation is everywhere and always 1 since, in a 
stationary state, they are balanced exactly by each other.

\section{SOLUTION WITH WINDS}

Although a fully advective solution has been automatically obtained in K00 
starting from the toroidal field of the form $\tilde{b}_{\varphi}\propto 
\xi^{-1}$, which is the consequence of $j_{\theta}=0$, we cannot use this 
relation here in obtaining a more general situation with non-zero $v_{\theta}$ 
and $j_{\theta}$. 
Instead, we start from the following form for the poloidal magnetic field, 
  \begin{equation}
   \tilde{b}_r(\xi) = B_0\ \xi^{-(3/2-n)},
  \end{equation}
where the parameter $n$ specifies the strength of a wind with $n=0$ 
corresponding to the no wind case. 

Substituting the above expression into equations (\ref{eqn:tRe}) and 
(\ref{eqn:Rfcns}), we obtain 
\begin{eqnarray}
   \tilde{b}_{\varphi}(\xi) &=& \Re_0 B_0\ \xi^{-(1-n)}, \\ 
   \Re_0 &=& \frac{3\pi(1-\alpha)^{1/2}(1-\beta)\sigma\Delta^2}{c^2} 
           \ l_{{\rm K}0}, 
\end{eqnarray}
and 
  \begin{equation}
    \tilde{b}_{\theta}(\xi) = \frac{2n+1}{4}\ \Delta B_0\ \xi^{-(3/2-n)},
  \end{equation}
respectively. 
Then, it follows from equation (\ref{eqn:Reomth}) that 
  \begin{equation}
   \tilde{p}(\xi) = \frac{\Re^2_0 B_0^2}{8\pi}\ \xi^{-2(1-n)}.
  \end{equation}
The subscript 0 is referred to each quantity at $r_0$. 

Combining the expressions for $\tilde{b}_r$ and $\tilde{b}_{\theta}$, 
we obtain from equation (\ref{eqn:Rindp}) 
  \begin{equation}
   \tilde{v}_r(\xi) = \frac{3(1-\alpha)^{1/2}(1-\beta)}{2n+1} 
    \ \frac{v_{{\rm K}0}}{\Re_0}\ \xi^{-1},
  \end{equation}
and further from this result and equation (\ref{eqn:Rmcns}), 
  \begin{equation}
   \tilde{v}_{\theta}(\xi) = \frac{6(1-\alpha)^{1/2}(1-\beta)n}{2n+1} 
    \ \frac{\Delta v_{{\rm K}0}}{\Re_0} \ \xi^{-1}. 
  \end{equation}
The density is calculated from equation (\ref{eqn:Reomph}) as 
  \begin{equation}
   \tilde{\rho}(\xi) = \frac{(2n+1)^2}{24\pi(1-\alpha)(1-\beta)}
    \ \frac{\Re^2_0 B_0^2}{v_{{\rm K}0}^2}\ \xi^{-(1-2n)}
  \end{equation}

With the above expressions for various quantities, we can determine 
first $\beta$ from the second IRAF condition (\ref{eqn:IRAF2}) and 
then $\alpha$ from the first condition (\ref{eqn:IRAF1}) as 
\begin{equation}
  \alpha = \frac{2}{3}(1-n), \quad \beta = \frac{2}{3}(1-n). 
\end{equation}
Therefore, the final forms of the above tentative expressions are 
\begin{equation}
  \Re(\xi) = \Re_0\ \xi^{1/2}, 
   \quad \Re_0 = \left( \frac{2n+1}{3} \right)^{3/2}
   \frac{3\pi\sigma\Delta^2 l_{{\rm K}0}}{c^2}
  \label{eqn:Re}
\end{equation}
  \begin{equation}
   \tilde{v}_r(\xi) = \sqrt{\frac{2n+1}{3}}\ \frac{v_{{\rm K}0}}{\Re_0}
   \ \xi^{-1},
  \end{equation}
  \begin{equation}
   \tilde{v}_{\theta}(\xi) = 2n\sqrt{\frac{2n+1}{3}} 
    \ \frac{\Delta v_{{\rm K}0}}{\Re_0} \ \xi^{-1},
  \end{equation}
  \begin{equation} 
    \tilde{v}_{\varphi}(\xi) 
    = \sqrt{\frac{2n+1}{3}}\ v_{{\rm K}0}\ \xi^{-1/2}, 
  \label{eqn:vph}
  \end{equation}
  \begin{equation}
   \tilde{\rho}(\xi) 
    = \frac{3\Re^2_0 B_0^2}{8\pi v_{{\rm K}0}^2}\ \xi^{-(1-2n)},
  \label{eqn:Rrho}
  \end{equation}
and the subsidiary equations yield 
  \begin{equation}
   \tilde{T}(\xi) = \frac{\bar{\mu}}{R}\ \frac{v_{{\rm K}0}^2}{3} 
   \ \xi^{-1},
  \end{equation}
  \begin{equation}
   \tilde{j}_r(\xi) = \left(\frac{c}{4\pi\Delta}\right) 
        \frac{\Re_0 B_0}{r_0}\ \xi^{-(2-n)},
  \end{equation}
  \begin{equation}
   \tilde{j}_{\theta}(\xi) = n \left(\frac{c}{4\pi}\right)
    \ \frac{\Re_0 B_0}{r_0}\ \xi^{-(2-n)},  
  \end{equation}
  \begin{equation}
   \tilde{j}_{\varphi}(\xi) = \left(\frac{c}{4\pi\Delta}\right) 
    \ \frac{B_0}{r_0}\ \xi^{-(5/2-n)}, 
  \end{equation}
  \begin{equation}
   \tilde{E}_r(\xi)=\tilde{E}_{\theta}(\xi)=\tilde{E}_{\varphi}(\xi)=0.
  \end{equation}

It turns out from the above expressions that the resistive IRAF solution 
has some characteristic features. 
First of all, temperature is independent of $n$ and always has the virial 
value. 
The radial powers of the all components of velocity are independent of $n$, 
while their coefficients depend on $n$. 
On the other hand, the powers of other quantities are dependent on $n$, 
while their coefficients are independent of $n$ except for 
$\tilde{b}_{\theta}$ and $\tilde{j}_{\theta}$. 
The vanishing of the electric field reflects one of our implicit 
assumption that the electric load such as the acceleration of a bipolar 
jet is completely neglected in obtaining the solution, and hence the total 
electric power available is spent in the accretion disk.

\section{ENERGY BUDGET}

Some global considerations of the physics associated with our accretion 
disk reveal more information about the present solution. 
First, from the definition of the disk's inner edge 
(i.e., $\Re(\xi_{\rm in}) = 1$) and equation (\ref{eqn:Re}), we obtain 
\begin{equation}
  \xi_{\rm in} = \Re_0^{-2}, 
\end{equation}
and the condition $r_{\rm in}\ll r_{\rm out}$ guarantees $\Re_0^{-2}\ll 1$. 

The total magnetic flux penetrating the disk surface can be regarded as 
being generated by the toroidal current in the disk, since the contribution 
of the seed field is negligible there. 
The whole flux closes beyond the outer edge, reducing the magnetic flux 
in that region. 
Then, the reduced amount of flux should appear in the narrow central region 
within the inner edge, with an enhanced strength. 
Approximating the mean intensity of the central magnetic field by 
$\tilde{b}_r(\xi_{\rm in})$, we have 
\begin{equation}
  \pi r_{\rm in}^2\ \tilde{b}_r(r_{\rm in}) 
  = \int_{r_{\rm in}}^{r_{\rm out}}\tilde{b}_{\theta}\ 2\pi r\ dr,
\end{equation}
which results in the relation 
\begin{equation}
  \Delta = \frac{\xi_{\rm in}^{n+1/2}}{1-\xi_{\rm in}^{n+1/2}} 
  \simeq \xi_{\rm in}^{n+1/2} = \Re_0^{-(2n+1)}.
\end{equation}
The approximate expression holds because $\xi_{\rm in}^{n+1/2}\ll 1$ 
(this point will be confirmed at the end of this section). 

Owing to the presence of vertical flows, the mass accretion rate becomes 
radius dependent like 
\begin{equation}
  \dot{M}(\xi) = -\int_{-\infty}^{\infty}2\pi\rho_r r^2\Delta\ d\eta 
  = \dot{M}_0\ \xi^{2n}, 
\end{equation}
where 
\begin{equation}
  \dot{M}_0 = \sqrt{\frac{2n+1}{3}}
  \ \frac{\Re_0^{-2n}B_0^{\ 2}r_0^{5/2}}{\sqrt{GM}}.
  \label{eqn:Md0}
\end{equation}
Solving equation (\ref{eqn:Md0}) for $r_0$, we obtain 
\begin{equation}
  r_0 = \left( \frac{3\Re_0^{4n}}{2n+1}\ \frac{GM\dot{M}_0^2}{B_0^{\ 4}} 
        \right)^{1/5}.
\end{equation}

It is quite easy to evaluate the Bernoulli sum for the present solution: 
\begin{equation}
  \mbox{Be}(\xi, \eta) \ \equiv \ 
    \frac{1}{2}v_{\varphi}^{\ 2} - \frac{GM}{r} + h 
  = \frac{n}{3}\ v_{{\rm K}0}^{\ 2}\xi^{-1} \equiv\ \mbox{\~Be}(\xi),
\end{equation}
where $h$ denotes the specific enthalpy and $h=(5/2)p/\rho$ for any 
ideal gas. 
The poloidal velocity does not appear in the kinetic energy term, since 
its contribution is small compared with that from the toroidal component 
by $\Re_0^{-2}$. 
The sum is not a constant along each stream line because the flow 
is dissipative. 
For non-zero $n$, its absolute value increases from zero at infinity 
as the radius decreases. 
It is apparent that, within the framework of the present solution, 
the sign of the Bernoulli sum is indeed a discriminator of wind 
generation since it directly reflects the sign of $n$: i.e., according 
to whether it is positive, zero, or negative we have up wind, no wind, 
or down wind, respectively. 
As discussed in \S 1, the appearance of non-zero sum may indicate 
the intervention of some other mechanism of energy transport than advection. 
This point becomes clear soon. 

The local energy budget is as follows. 
The Joule heating rate is calculated as 
\begin{eqnarray}
  q_{\rm J}^{+}(\xi, \eta) &=& \frac{{\bf j}^2}{\sigma}\sim
    \frac{j_r^2}{\sigma} \nonumber \\
  &=& \frac{2n+1}{16\pi}\ \Re_0^{2n+1}\ \frac{GM\dot{M}_0}{r_0^4}
    \ \xi^{-2(2-n)}\mbox{sech}^4\eta, 
\end{eqnarray}
where $c^2/\sigma\Delta^2$ and $B_0^2$ have been eliminated by the aid 
of equations (\ref{eqn:Re}) and (\ref{eqn:Md0}). 
The advection cooling, on the other hand, is 
\begin{eqnarray}
  q_{\rm adv}^{-}(\xi, \eta) &\equiv& 
   \mbox{div}(h\rho{\bf v}) - ({\bf v}\cdot\mbox{grad})p \nonumber \\
   &=& \frac{1}{r^2}\frac{\partial}{\partial r}(r^2h\rho v_r) 
    + \frac{1}{r\Delta}\frac{\partial}{\partial\eta}(h\rho v_{\theta}) 
    - v_r\frac{\partial p}{\partial r} 
    - \frac{v_{\theta}}{r\Delta} \frac{\partial p}{\partial\eta} \nonumber \\
  &=& \frac{4n+1}{16\pi}\ \Re_0^{2n+1}\ \frac{GM\dot{M}_0}{r_0^4} 
    \ \xi^{-2(2-n)}\ \mbox{sech}^4\eta, 
\end{eqnarray}
where the last term in the second line should be dropped on account of 
the assumption v). 

It is evident, therefore, that the advective cooling is in general not 
balanced by the Joule heating alone. 
This means that there must be some additional heating that have not been 
mentioned explicitly, so that the energy balance in a stationary state 
should in fact be 
\begin{equation}
  q^+_{\rm J} + q^+_{\rm add} = q^-_{\rm adv} 
\end{equation}
since radiation cooling is negligible ($q^-_{\rm adv}\gg q^-_{\rm rad}$) 
also in the present solution. 
From the above expressions for $q_{\rm J}^+$ and $q_{\rm adv}^-$, we have 
\begin{equation}
  q^+_{\rm add} = \frac{n}{8\pi}\ \Re_0^{2n+1}
    \ \frac{GM\dot{M}_0}{r_0^4}\ \xi^{-2(2-n)}\mbox{sech}^4\eta. 
\end{equation}
This is what has been suggested just above by the appearance of a 
non-zero Bernoulli sum. 

Plausible examples of such additional heating mechanisms may be viscous 
heating ($q^+_{\rm vis}$, as is popular in the standard and viscous ADAF 
theories), convection heating ($q^+_{\rm conv}$, e.g., Narayan, 
Igumenshchev, \& Abramowicz 2000; Quataert \& Gruzinov, 2000), 
conduction heating ($q^+_{\rm cond}$) and so on. 
Among these examples, the first one should always be positive, while 
the others may be negative as well. 
The above result for $q^+_{\rm add}$ means that we have up wind, no 
wind or down wind according to whether the additional heating rate is 
positive, zero or negative, respectively. 
In order to keep the global energy budget in a disk, the role of these 
additional mechanisms should be the redistribution of locally available 
gravitational energy. 

Finally, we derive the restriction on the rage of the wind parameter 
$n$. 
Upper limit for $n$ is derived from the radial dependence of $\rho$ 
in equation (\ref{eqn:Rrho}). 
Since the density should decrease outward, we have $n<1/2$. 
The most stringent lower limit for $n$ is obtained if we require that 
the advective cooling should be really cooling (i.e., $-1/4<n$ when 
$q^-_{\rm adv}>0$). 
This is equivalent to the requirement that entropy should decrease 
outward, because in a stationary state we can show the equality 
$q^-_{\rm adv} = \rho T({\bf v}\cdot\mbox{grad})s$, where $s$ is the 
specific entropy. 
Thus, we have obtained 
\begin{equation}
  -\frac{1}{4} < n < \frac{1}{2}. 
\end{equation}
This inequality guarantees $\xi_{\rm in}^{n+1/2}\ll 1$ as far as 
$\xi_{\rm in}\ll 1$.

\section{SUMMARY AND CONCLUSION}

We have extended our former analytic solution for ADAF in a global 
magnetic field to include winds from the disk surface. 
The solution contains one parameter $n$ ($-1/4<n<1/2$) that specifies 
the sense and strength of the winds. 
According to whether $n$ is positive, zero or negative, we have up wind, 
no wind or down wind, respectively, with the wind strength increased 
according to its absolute value. 

The following physically different statements are all equivalent 
mathematically, and represent the criterion for the generation of up winds 
(i.e., $n>0$). \\
(1) Poloidal magnetic field in the disk decreases with radius slower 
than $r^{-3/2}$. \\
(2) Rotational velocity of the disk is sufficiently large: i.e., 
$v_{\varphi}>(1/\sqrt{3})v_{\rm K}$. \\
(3) Pressure gradient is sufficiently small compare with gravity: 
i.e., $\alpha<2/3$. \\
(4) The Bernoulli sum in the disk is positive. \\
(5) There are other mechanisms of heating that has not been treated 
explicitly in the present consideration. 

Judging from the radial dependence of the poloidal magnetic field, 
the converging flow of Kaburaki (1978) corresponds to the extreme 
case of $n=-1/4$ (down wind). 
Although the poloidal field of the Blandford-Payne solution (1982) 
formally corresponds to the case of $n=1/4$ (up wind), the correspondence 
should not be taken so seriously because their self-similar solution is 
somewhat artificial and does not belong to the same class as the solution 
derived here. 
The second statement above suggests that the winds are centrifugally 
driven, and the third, that strong winds apt to appear in high pressure 
environments as discussed by Fabian \& Rees (1995). 

The fourth and fifth statements are closely related. 
The Bernoulli sum works as a discriminator of wind generation, at least 
within the framework of present treatment. 
The appearance of non-zero sum seems to indicate the intervention of some 
other mechanisms of energy transport than advection. 
These additional processes act to redistribute the locally available 
gravitational energy, within a disk. 
It may be heat conduction or convection that carries thermal energy from 
inner regions to outer . 
Another important possibility may be the viscosity that has been completely 
neglected in this paper for simplicity, but is generally known to have such 
a redistribution effect (e.g., Frank, King, \& Raine 1992). 
In either case, a more satisfactory solution should be obtained by including 
each process explicitly in the treatment, and it is of course beyond the 
scope of this paper. 

The wind velocity described by our solution is sub-critical even if 
it originates from an inner portion of the accretion disk, since 
$\tilde{v}_{\theta}/v_{\rm K}\sim \Delta/\Re(r)$ ($\Delta\ll 1$ and 
$\Re^{-1}(r)<1$), and the sound and Alfv\'{e}n velocities are always 
of the order of the Kepler velocity, i.e., $V_{\rm S}(r) \simeq 
V_{\rm A}(r) = \sqrt{2/3}\ v_{\rm K}(r)$ where $V_{\rm S}^2\equiv 
d\tilde{p}/d\tilde{\rho}$ and $V_{\rm A}^2 \equiv 
\tilde{b}_{\varphi}^2/4\pi\tilde{\rho}$. 
Therefore, the simple collimation of this type of wind from the surface 
of an accretion disk cannot explain the formation of a relativistic jet. 
The generation cite of the latter should probably be attributed to 
the central region within the inner edge of disk. 
In spite of low velocities expected for winds from the disk surfaces, 
their effects on the accretion process can be appreciable as seen from 
the resulting radius-dependent mass accretion rates.

\end{document}